\documentclass[zpreprint,zbstdefault]{./LaTeX/zeus/zeus_paper}
\usepackage[english]{babel}

\newcommand{\ZcoosysB}{%
The ZEUS coordinate system is a right-handed Cartesian system, with the $Z$
axis pointing in the proton beam direction, referred to as the ``forward
direction'', and the $X$ axis pointing  towards the centre of HERA.
The coordinate origin is at the nominal interaction point.\xspace}

\newcommand{\ZcoosysfnB}{\footnote{\ZcoosysB}}

\newcommand{\Zdetdesc}{%
A detailed description of the ZEUS detector can be found 
elsewhere~\cite{zeus:1993:bluebook}. A brief outline of the 
components that are most relevant for this analysis is given
below.\xspace}
\newcommand{\Zctddesc}[1]{%
Charged particles were tracked in the central tracking detector (CTD)~\citeCTD,
which operated in a magnetic field of $1.43\Tesla$ provided by a thin 
superconducting coil. The CTD consisted of 72~cylindrical drift chamber 
layers, organised in 9~superlayers covering the polar-angle#1 region 
\mbox{$15^\circ<\theta<164^\circ$}. The transverse-momentum resolution for
full-length tracks was $\sigma(p_T)/p_T=0.0058p_T\oplus0.0065\oplus0.0014/p_T$,
with $p_T$ in $\Gev$.}
\newcommand{\Zcaldesc}{%
The high-resolution uranium--scintillator calorimeter (CAL)~\citeCAL consisted 
of three parts: the forward (FCAL), the barrel (BCAL) and the rear (RCAL)
calorimeters. Each part was subdivided transversely into towers and
longitudinally into one electromagnetic section (EMC) and either one (in RCAL)
or two (in BCAL and FCAL) hadronic sections (HAC). The smallest subdivision of
the calorimeter is called a cell.  The CAL energy resolutions, as measured under
test-beam conditions, were $\sigma(E)/E=0.18/\sqrt{E}$ for electrons and
$\sigma(E)/E=0.35/\sqrt{E}$ for hadrons, with $E$ in $\Gev$.}
\chardef\usc=95
\chardef\til=126
\catcode`\@=11 % @ signs are now treated as letters
\DeclareRobustCommand\xdotspace{\futurelet\@let@token\@xdotspace}
\def\@xdotspace{%
  \ifx\@let@token.\else
  \ifx\@let@token\bgroup.\else
  \ifx\@let@token\egroup.\else
  \ifx\@let@token\/.\else
  \ifx\@let@token\ .\else
  \ifx\@let@token~.\else
  \ifx\@let@token!.\else
  \ifx\@let@token,.\else
  \ifx\@let@token:.\else
  \ifx\@let@token;.\else
  \ifx\@let@token?.\else
  \ifx\@let@token/.\else
  \ifx\@let@token'.\else
  \ifx\@let@token).\else
  \ifx\@let@token-.\else
  \ifx\@let@token\@xobeysp.\else
  \ifx\@let@token\space.\else
  \ifx\@let@token\@sptoken.\else
   .\space
   \fi\fi\fi\fi\fi\fi\fi\fi\fi\fi\fi\fi\fi\fi\fi\fi\fi\fi}
\catcode`\@=12 % @ signs are no longer letters

\newcommand{\stru}[2]{%
   \relax\ifmmode\hbox{\vrule height#1 depth#2 width0pt}%
   \else\vrule height#1 depth#2 width0pt\fi}
\newcommand{\Ronum}[1]{\uppercase\expandafter{\romannumeral#1}}
\newcommand{\ronum}[1]{\expandafter{\romannumeral#1}}
\DeclareRobustCommand{\LaTeXZ}{%
  \LaTeX\kern-.05em4\kern-.1em
  {\raisebox{-0.2ex}{$\scriptstyle\text{ZEUS}$}}\xspace}
\newcommand{\eq}[1]{(\ref{eq-#1})}

\newcommand{\fig}[1]{Fig.~\ref{fig-#1}}

\newcommand{\tab}[1]{Table~\ref{tab-#1}}

\DeclareMathAlphabet{\mathbf}{OT1}{cmr}{bx}{sl}
\newcommand{\eVdist}{\kern-0.06667em}

\newcommand{\Gev}{{\text{Ge}\eVdist\text{V\/}}}

\newcommand{\mev}{{\,\text{Me}\eVdist\text{V\/}}}
\newcommand{\gev}{{\,\text{Ge}\eVdist\text{V\/}}}

\newcommand{\nb}{\,\text{nb}}
\newcommand{\pb}{\,\text{pb}}

\newcommand{\met}{\,\text{m}}

\newcommand{\cm}{\,\text{cm}}

\newcommand{\Tesla}{\,\text{T}}

\newcommand{\slashfrac}[2]{%
  \raisebox{0.5ex}{\ensuremath #1}\kern-0.12em/\kern-0.08em
  \raisebox{-.8ex}{\ensuremath #2}}
\newcommand{\sqr}[3]{%
    {\vcenter{\hrule height.#3ex\hbox{\vrule width.#2ex height#1ex
     \kern#1ex\vrule width.#3ex}\hrule height.#2ex}}}

\catcode`\@=11 % @ signs are now treated as letters
\newcommand{\parenbar}{\mathpalette\p@renb@r}
\def\p@renb@r#1#2{\vbox{%
  \ifx#1\scriptscriptstyle \dimen@.7em\dimen@ii.2em\else
  \ifx#1\scriptstyle \dimen@.8em\dimen@ii.25em\else
  \dimen@1em\dimen@ii.4em\fi\fi \offinterlineskip
  \ialign{\hfill##\hfill\cr
    \vbox{\hrule width\dimen@ii}\cr
    \noalign{\vskip-.3ex}%
    \hbox to\dimen@{$\mathchar300\hfil\mathchar301$}\cr
    \noalign{\vskip-.3ex}%
    $#1#2$\cr}}}
\catcode`\@=12 % @ signs are no longer letters

\newcommand{\IP}{{\rm I$\kern-0.01667em$P}\xspace}

\mathchardef\qsm=63
\mathchardef\pls=43
\mathchardef\mns=512
\mathchardef\plm=518
\mathchardef\eql=61
\mathchardef\smallleft=300
\mathchardef\smallright=301
\mathchardef\les=316
\mathchardef\gre=318
\mathchardef\leq=532
\mathchardef\grq=533
\catcode`\@=11 % @ signs are now treated as letters
\newcounter{pict@width}
\newcounter{pict@height}
\newlength{\pict@scale}
\newcommand{\psfigadd}[4]{%
\setcounter{pict@width}{1*\ratio{#2+\pict@scale/2}{\pict@scale}}
\setcounter{pict@height}{1*\ratio{#3+\pict@scale/2}{\pict@scale}}
\setlength{\unitlength}{\pict@scale}
\hbox to #2{\hspace{-\fill}\begin{picture}(\thepict@width,\thepict@height)
\put(0,0){\psfig{figure=#1,width=#2,height=#3,clip=}}
\SetScale{0.283466457}
\SetWidth{1.763889}
{#4}
\end{picture}}
}
\newcounter{pict@widthfst}
\newcounter{pict@widthscd}
\newcounter{pict@widthtot}
\newcommand{\psfigaddtwo}[7]{%
\setcounter{pict@widthfst}{1*\ratio{#2+\pict@scale/2}{\pict@scale}}
\setcounter{pict@widthscd}{1*\ratio{#2+#4+\pict@scale/2}{\pict@scale}}
\setcounter{pict@widthtot}{1*\ratio{#2+#4+#6+\pict@scale/2}{\pict@scale}}
\setcounter{pict@height}{1*\ratio{#3+\pict@scale/2}{\pict@scale}}
\setlength{\unitlength}{\pict@scale}
\hbox{\hspace{-\fill}\begin{picture}(\thepict@widthtot,\thepict@height)
\put(0,0){\psfig{figure=#1,width=#2,height=#3,clip=}}
\put(\thepict@widthscd,0){\psfig{figure=#5,width=#6,height=#3,clip=}}
\SetScale{0.283466457}
\SetWidth{1.763889}
{#7}
\end{picture}}
}
\newcommand{\psfigror}[4]{%
\setcounter{pict@width}{1*\ratio{#2+\pict@scale/2}{\pict@scale}}
\setcounter{pict@height}{1*\ratio{#3+\pict@scale/2}{\pict@scale}}
\setlength{\unitlength}{\pict@scale}
\hbox{\begin{picture}(\thepict@width,\thepict@height)
\put(0,\thepict@height){\psfig{figure=#1,width=#3,height=#2,clip=,angle=270}}
\SetScale{0.283466457}
\SetWidth{1.763889}
{#4}
\end{picture}}
}
\newcommand{\psfigrol}[4]{%
\setcounter{pict@width}{1*\ratio{#2+\pict@scale/2}{\pict@scale}}
\setcounter{pict@height}{1*\ratio{#3+\pict@scale/2}{\pict@scale}}
\setlength{\unitlength}{\pict@scale}
\hbox{\begin{picture}(\thepict@width,\thepict@height)
\put(0,0){\psfig{figure=#1,width=#3,height=#2,clip=,angle=90}}
\SetScale{0.283466457}
\SetWidth{1.763889}
{#4}
\end{picture}}
}
\catcode`\@=12 % @ signs are no longer letters
\newlength\listtextwidth

\catcode`\@=11 % @ signs are now treated as letters
\newlength{\@tabfninsert}
\newlength{\@tabfnwidth}
\newcommand{\tabfootnote}[2]{%
  \setlength{\@tabfninsert}{0.8em}
  \setlength{\@tabfnwidth}{\textwidth}
  \addtolength{\@tabfnwidth}{-\@tabfninsert}
  \addtolength{\@tabfnwidth}{-0.4em}
  \noindent\makebox[\@tabfninsert][r]{\footnotesize$^{#1}$\hfil}\hfill%
  \parbox[t]{\@tabfnwidth}{\footnotesize #2\hfill}}
\catcode`\@=12 % @ signs are no longer letters

\def\citeCTD{{\cite{%
nim:a279:290,*npps:b32:181,*nim:a338:254%
}}\xspace}
\def\citeCAL{{\cite{%
nim:a309:77,*nim:a309:101,*nim:a321:356,*nim:a336:23%
}}\xspace}
\begin{document}
%------------------------------------------------------------------------------
%       Title sheet
%------------------------------------------------------------------------------
\prepnum{DESY--09--137}

\title{
Measurement of \boldmath$J/\psi$ 
photoproduction  
at large momentum transfer at HERA
}                                                       
                    
\author{ZEUS Collaboration}
\draftversion{}
\date{September, 2009}

\abstract{
The proton-dissociative diffractive photoproduction of $J/\psi$ mesons has
been studied in $ep$ collisions with the ZEUS detector at HERA using an
integrated luminosity of 112 $\rm pb^{-1}$. The cross section is presented
as a function of the  photon-proton centre-of-mass energy 
and of the squared four-momentum transfer at the proton vertex. The
results are compared to perturbative QCD calculations.

}

\makezeustitle

\def\3{\ss}                                                                                        
\pagenumbering{Roman}                                                                              
\begin{center}                                                                                     
{                      \Large  The ZEUS Collaboration              }                               
\end{center}                                                                                       
  S.~Chekanov,                                                                                     
  M.~Derrick,                                                                                      
  S.~Magill,                                                                                       
  B.~Musgrave,                                                                                     
  D.~Nicholass$^{   1}$,                                                                           
  \mbox{J.~Repond},                                                                                
  R.~Yoshida\\                                                                                     
 {\it Argonne National Laboratory, Argonne, Illinois 60439-4815, USA}~$^{n}$                       
\par \filbreak                                                                                     
  M.C.K.~Mattingly \\                                                                              
 {\it Andrews University, Berrien Springs, Michigan 49104-0380, USA}                               
\par \filbreak                                                                                     
  P.~Antonioli,                                                                                    
  G.~Bari,                                                                                         
  L.~Bellagamba,                                                                                   
  D.~Boscherini,                                                                                   
  A.~Bruni,                                                                                        
  G.~Bruni,                                                                                        
  F.~Cindolo,                                                                                      
  M.~Corradi,                                                                                      
\mbox{G.~Iacobucci},                                                                               
  A.~Margotti,                                                                                     
  R.~Nania,                                                                                        
  A.~Polini\\                                                                                      
  {\it INFN Bologna, Bologna, Italy}~$^{e}$
% ...
\newpage                                                                                           
$^{\    1}$ also affiliated with University College London,                                        
United Kingdom\\                                                                                   
% ...
%                                                                                                  
% \par         % if index listing & table fit to 1 page, put gap here                              
\newpage   % alternatively: go to newpage, if page is too small                                    
                                                           %                                       
% \institute_references_start    % do not touch or move this line !                                
                                                           %                                       
\begin{tabular}[h]{rp{14cm}}                                                                       
$^{a}$ &  supported by the Natural Sciences and Engineering Research Council of Canada (NSERC) \\  
$^{b}$ &  supported by the German Federal Ministry for Education and Research (BMBF), under        
          contract Nos. 05 HZ6PDA, 05 HZ6GUA, 05 HZ6VFA and 05 HZ4KHA\\                            
$^{c}$ &  supported in part by the MINERVA Gesellschaft f\"ur Forschung GmbH, the Israel Science   
          Foundation (grant No. 293/02-11.2) and the US-Israel Binational Science Foundation \\    
$^{d}$ &  supported by the Israel Science Foundation\\                                             
$^{e}$ &  supported by the Italian National Institute for Nuclear Physics (INFN) \\                
$^{f}$ &  supported by the Japanese Ministry of Education, Culture, Sports, Science and Technology 
          (MEXT) and its grants for Scientific Research\\                                          
$^{g}$ &  supported by the Korean Ministry of Education and Korea Science and Engineering          
          Foundation\\                                                                             
$^{h}$ &  supported by the Netherlands Foundation for Research on Matter (FOM)\\                   
$^{i}$ &  supported by the Polish State Committee for Scientific Research, project No.             
          DESY/256/2006 - 154/DES/2006/03\\                                                        
$^{j}$ &  partially supported by the German Federal Ministry for Education and Research (BMBF)\\   
$^{k}$ &  supported by RF Presidential grant N 1456.2008.2 for the leading                         
          scientific schools and by the Russian Ministry of Education and Science through its      
          grant for Scientific Research on High Energy Physics\\                                   
$^{l}$ &  supported by the Spanish Ministry of Education and Science through funds provided by     
          CICYT\\                                                                                  
$^{m}$ &  supported by the Science and Technology Facilities Council, UK\\                         
$^{n}$ &  supported by the US Department of Energy\\                                               
$^{o}$ &  supported by the US National Science Foundation. Any opinion,                            
findings and conclusions or recommendations expressed in this material                             
are those of the authors and do not necessarily reflect the views of the                           
National Science Foundation.\\                                                                     
$^{p}$ &  supported by the Polish Ministry of Science and Higher Education                         
as a scientific project (2009-2010)\\                                                              
$^{q}$ &  supported by FNRS and its associated funds (IISN and FRIA) and by an Inter-University    
          Attraction Poles Programme subsidised by the Belgian Federal Science Policy Office\\     
$^{r}$ &  supported by an FRGS grant from the Malaysian government\\                               
\end{tabular}                                                                                      
                                                           %                                       
% \institute_references_end     % do not touch or move this line !                                 
                                                           %                                       

%------------------------------------------------------------------------------
%       Text
%------------------------------------------------------------------------------
\pagenumbering{arabic} 
\pagestyle{plain}
% ----------------------------------------------------------------------------
%       Introduction
% ----------------------------------------------------------------------------
\newcommand {\pom} {I\!\!P}
\newcommand {\pomsub} {{\scriptscriptstyle \pom}}
\newcommand {\apom} {\alpha_{\pomsub}}
\newcommand {\aprime} {\alpha^\prime_\pomsub}

\section{Introduction}
\label{sec-int}

Photoproduction of vector mesons (VMs) is usually thought of as a
process where the photon fluctuates into a $q \bar{q}$ state, which
then interacts with the proton and becomes a VM. If the spatial
configuration of the $q \bar{q}$ state is large, its interaction with
the proton is soft in nature and is usually described by Regge
theory~\cite{collins:1977:regge} together with the vector dominance
model~\cite{anphy:11:1,*prl:22:981}. This applies to exclusive
photoproduction of the light VMs $\rho, \omega$ and $\phi$ 
(see Ivanov, Nikolaev and Savin \cite{vmreview} for a recent review). 
For heavy VMs, the $q\bar{q}$ pair is squeezed into a small configuration and 
 perturbative QCD (pQCD)~\cite{hfs} can be applied. In exclusive
photoproduction of $J/\psi$, $\gamma
\ p \to J/\psi \ p$, the mass of the $J/\psi$ provides a hard scale at the
photon vertex and the 
small-size  $q\bar{q}$  pair interacts through a
two-gluon ladder with partons in the proton. If the
four-momentum-transfer squared at the proton vertex is small,
$|t| \lesssim$ 1\gev$^2$, and the proton stays intact, 
the cross section is predicted to fall exponentially with $|t|$.
 
When $|t|$ increases, $|t| > 1\gev^2$, the dominant process is
that where the proton dissociates into a low-mass nucleon state
$Y$,
\begin{equation}
\gamma \ p \to J/\psi \ Y.
\label{eq:jpsi}
\end{equation}
At large $|t|$ values, the cross section is expected to have a
power-law decrease with $|t|$~\cite{fr,bartels,forshaw,enberg}. In
addition,  $J/\psi$ photoproduction at large $|t|$ is a two-scale
process in which the large mass of the heavy VM is the hard scale at
the photon vertex and $t$ is the hard scale at the proton vertex. At
high photon-proton centre-of-mass energies, $W$, this process should
be sensitive to BFKL~\cite{jetp:44:443,*jetp:45:199,*sovjnp:28:822} dynamics.

This paper contains results for the kinematic range $30 < W <
160\gev$ and $ 2 < |t| < 20\gev^2$, which is larger than for the
previous ZEUS measurement~\cite{epj:c26:389}. The sample under study also
represents more than a five-fold increase in integrated luminosity.

% ----------------------------------------------------------------------------
%       Experimental set-up
% ----------------------------------------------------------------------------
\section{Experimental set-up}
\label{sec-exp}

This analysis is based on  data collected with the ZEUS detector at
HERA in 1996$-$2000. In those years HERA operated with an electron\footnote{Electrons and positrons are both referred to as electrons in this paper.}
beam energy of $27.5\gev$ and a proton beam energy, $E_p$, of $820\gev$
(1996$-$1997) and $920\gev$ (1998$-$2000).  The data sample corresponds
to an integrated luminosity of $112\pb^{-1}$, $36\pb^{-1}$ with $E_p=820\gev$ and $76\pb^{-1}$ with $E_p=920\gev$.

\Zdetdesc

\Zctddesc\ZcoosysfnB~ \Zcaldesc

The muon system \cite{nim:a333:342} consisted of tracking detectors
(forward, barrel and rear muon chambers: FMUON, B/RMUON), which were
placed inside and outside a magnetised iron yoke surrounding the CAL.
The inner chambers, F/B/RMUI, covered the polar angles from $10^\circ$
to $34^\circ$, from $34^\circ$ to $135^\circ$ and from $135^\circ$ to
$171^\circ$, respectively.

The luminosity was determined from the rate of the bremsstrahlung
process $ep\rightarrow e\gamma p$, where the photon was measured by a
lead-scintillator calorimeter~\cite{acpp:b32:2025} located at $Z=-107 \met$.

%----------------------------------------
\section {Kinematics and reconstruction}
%----------------------------------------

\begin{figure}[h]
\begin{center}
\includegraphics[bb = 60 325 440 580, scale=0.7]{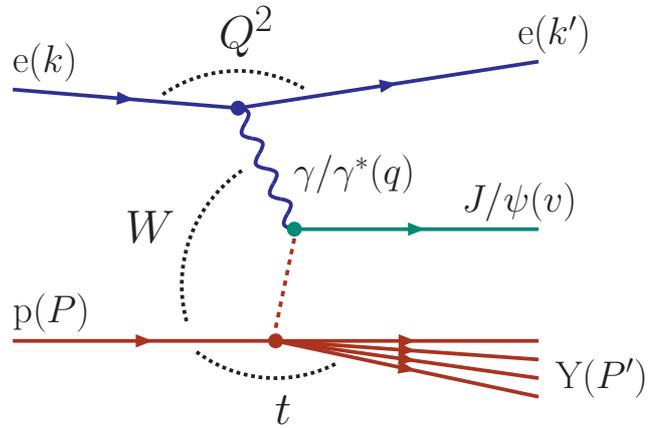}
\end{center}
\caption{Schematic diagram of proton-dissociative $J/\psi$ production in $ep$
interactions, $ep \rightarrow e J/\psi~Y$.
\label{fig-fig1}
}
\end{figure}

The proton-dissociative $J/\psi$ production process in $e p$
interactions,
\begin{displaymath}
e(k)p(P)\rightarrow e(k')J/\psi(v)Y(P'),
\end{displaymath}
is illustrated in \fig{fig1}.  The signature of these events
consists of two oppositely charged muons from the $J/\psi$ decay and
of the remnant of the dissociated proton. In the case of photoproduction,
the beam electron is scattered at small angles and escapes undetected
down the beampipe.

The variables $k$, $k'$, $P$, $P'$ and $v$ are the four-momenta of the
incident electron, scattered electron, incident proton, diffractive
nucleonic system $Y$ and $J/\psi$, respectively. The four-momentum of
the exchanged photon is denoted by $q$. The kinematic variables are
the following:
\begin{itemize}
\item $Q^2 = -q^2 = -(k-k')^2 $, the negative squared four-momentum  of the exchanged 
photon;

\item $W^2 = (q+P)^2$, the squared centre-of-mass energy of the photon-proton system;
\item $t = (P-P')^2 = (q-v)^2$, the squared four-momentum transfer at the proton vertex;
\item $y=(P\cdot q)/(P\cdot k)$, the fraction of the electron energy transferred 
to the photon in the rest frame of the proton;
\item $z=(P\cdot v)/(P\cdot q)$,  the event inelasticity, i.e. the fraction of the 
virtual photon energy transferred to the $J/\psi$ in the proton rest
frame.
\end{itemize}

The dissociated proton either escapes undetected down the beampipe or
deposits only a part of its energy in the CAL and hence 
the mass of the proton remnant, $M_Y$, cannot be measured precisely.
However, $M_Y$ is  related to other kinematic variables through $M^2_Y = W^2 (1 - z) - |t|$.

\begin{figure}[h]
\begin{center}
\includegraphics[ scale=0.7]{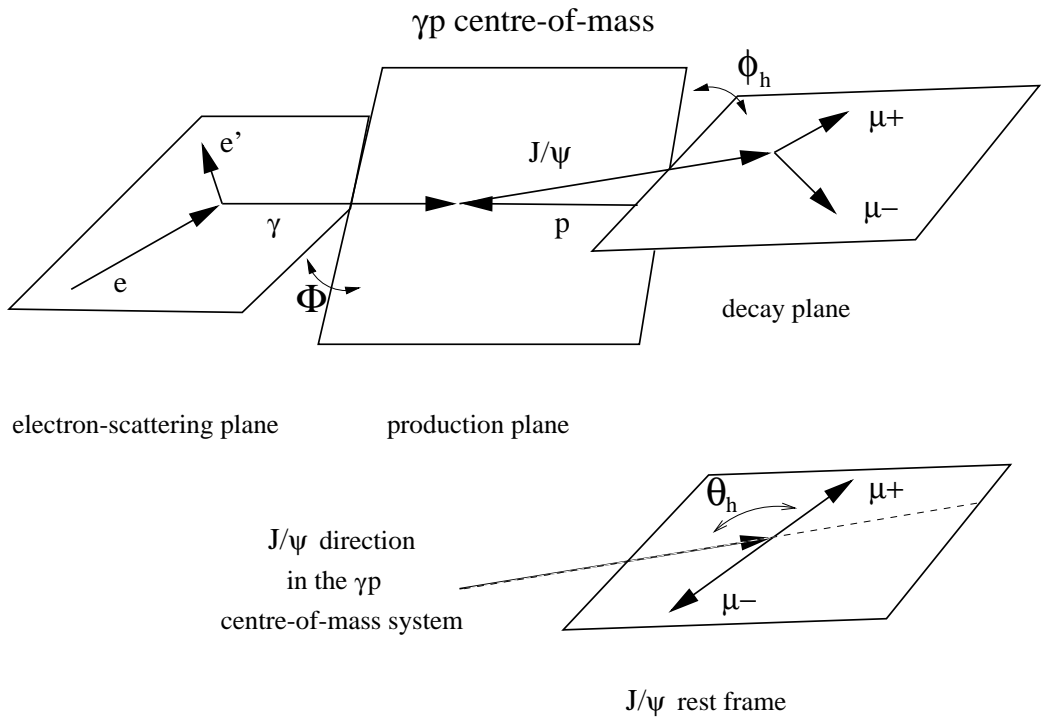}
\end{center}
\caption{Angles used to analyse the helicity states of the 
$J/\psi$, see text.}
\label{fig-fig2}
\end{figure}

The following angles are used to describe the decay of the $J/\psi$
(see \fig{fig2}):
\begin{itemize}
\item $\Phi$, the angle between the 
electron-scattering plane and the vector-meson plane, in the
photon-proton centre-of-mass frame;
\item $\theta_h$ and $\phi_h$, the polar and azimuthal angles of the 
positively-charged decay particle in the helicity frame. Here, the helicity frame
is the $J/\psi$ rest frame and the quantisation axis is the meson
direction in the photon-proton centre-of-mass system. The polar angle,
$\theta_h$, is defined as the angle between the direction of the
positively charged decay particle and the quantisation axis. The
azimuthal angle, $\phi_h$, is the angle between the decay plane and
the vector-meson production plane.
\end{itemize}

In this study, photoproduction is characterised by the non-observation of the scattered electron.
Thus, $Q^2$ ranges from the kinematic minimum, $Q^2_{\text
{min}}=m^2_ey^2/(1-y)\approx 10^{-7}\gev^2$, where $m_e$ is the
electron mass, up to $Q^2_{\text {max}}\approx 1\gev^2$, the value at
which the scattered electron becomes observable in the CAL.  Since
the mean $Q^2$ is small, $\langle Q^2 \rangle \approx 5\cdot 10^{-5}\gev^2$,
 it was neglected in the reconstruction of the other kinematic
variables.

The variable $t$ can be expressed as $t\approx -p^2_T$, where $p_T$ is
the transverse momentum of the produced vector meson in the laboratory
frame.  The variable $W$ is calculated as $W^2\approx
2E_p(E-p_Z)_{J/\psi}$, where $E$ is the energy and $p_Z$ is the longitudinal momentum of
the vector meson.  The quantities $(E-p_Z)_{J/\psi}$ and $t$ were
reconstructed using only the measured momenta of the VM muon decay
particles.

The inelasticity $z$  was computed from $z=(E-p_Z)_{J/\psi}/\sum(E-p_Z)$,
where $\sum(E-p_Z)= (E-p_Z)_{J/\psi}+\sum(E-p_Z)_{\rm had}$ and
$\sum(E-p_Z)_{\rm had}$ is reconstructed by summing over all the CAL
energy deposits (larger than 300 MeV) not associated with the  $J/\psi$
candidate. 

%-----------------------------
\section {Event selection}
\label{sec-rec}
%-----------------------------

The events were selected online by the ZEUS three-level trigger
system \cite{zeus:1993:bluebook,uproc:chep:1992:222}.  The events were
required to have at least one track in the CTD. At least one track had
to point towards a CAL energy deposit compatible with a minimum
ionising particle  as well as a signal in the inner muon
chambers.

The following was required offline:
\begin{itemize}
\item no scattered electron observed;
\item two tracks with opposite charge pointing to a primary vertex with 
$|Z_{\rm vertex}|< 50 \cm$;
\item both tracks well reconstructed, i.e. traversing at least three 
superlayers in the CTD, including the innermost layer;
\item each track associated with a distinct CAL energy deposit  within a
radius of $30\cm$;
\item azimuthal angle between the two 
tracks associated with the two muon candidates less than $174^\circ$ in order to reject cosmic-ray events;
\item  invariant mass of the two tracks, which were assigned a $\mu$ mass, in 
the range $2.6<M_{\mu\mu}<3.5\gev$. 
\end{itemize}

Events were required to be in a kinematic range where the properties
of the final state particles were properly measured and the acceptance
was well defined. This was satisfied for $2<|t|<20\gev^2$ and
$30<W<160\gev$. The cut of  $|t|>2\gev^2$ also significantly reduced the
background from the exclusive process.
A cut of $z>0.95$ was applied to suppress non-diffractive background. 
This cut also restricted the invariant mass of the $Y$ system
to $M_Y<30\gev$.

The energy range $30<W<40\gev$ was mainly populated by events  triggered by 
the FMUON detector, while the range $40<W<160\gev$ was dominated by
B/RMUON-triggered events. The FMUON-triggered sample was limited to the
data collected in 1996$-$1997 and covered the $|t|$ region up to
$10\gev^2$.

After this selection procedure the number of observed di-muon events was $2817$.

%%%%%%%%%%%%%%%%%%%%%%%%%%%%%%%%%%%%%%%%%%%%%%%55
\section{Theoretical predictions}
%%%%%%%%%%%%%%%%%%%%%%%%%%%%%%%%%%%%%%%%%%%%%%%%%

The reaction $\gamma \ p \to J/\psi \ Y$ can be viewed as a three-step
process. The photon fluctuates into a $q\bar{q}$ pair that scatters
off a single parton in the proton by the exchange of a colour
singlet. The scattered $q\bar{q}$ pair becomes a $J/\psi$ and the
struck parton and the proton remnant together  fragment into the system $Y$.
 In  lowest-order QCD the
colour singlet exchanges a pair of gluons. In the
leading logarithmic (LL) approximation, the process is described by
the effective exchange of a gluon ladder.

As stated in the introduction, the process under study has two scales.
 At the photon vertex, where the photon fluctuates into a
$q\bar{q}$ pair, the size is fixed and determined by the $J/\psi$
mass. The second scale, $|t|$, controls the size of the
system which emits the gluon ladder. 

In the region where the scale  $|t|$ is smaller than $M_{J/\psi}^2$
($2<|t|<10\gev^2$), the momenta on the gluon ladder are still
expected to be ordered and thus a
DGLAP~\cite{dglap2,*dglap4,*dglap5,*dglap7}
approach is appropriate. A calculation in this kinematic region has
been carried out by Gotsman, Levin, Maor and Naftali
(GLMN)~\cite{glmn},  using their screening correction formalism and
evolving the gluon in a LL DGLAP mechanism. 

As $|t|$ increases, the BFKL mechanism is expected to dominate. The
first LL BFKL calculations~\cite{fr,bartels,forshaw} were made using the
Mueller-Tang (MT) approximation~\cite{mt}, which is only good for very
large rapidity intervals.  Enberg, Motyka and Poludniowski
(EMP)~\cite{enberg} do not use the MT approximation and provide a
complete analytical solution in LL for the case of  heavy quarks (the case for
any quark mass is discussed elsewhere~\cite{enberg1,enberg2}). They use two
 different values of $\alpha_S$  as the pre-factor of the cross section and as the
coupling relevant for the BFKL ladder. Enberg at al.~\cite{enberg}
also present results of a non-leading (nonL) BFKL calculation.

In addition,  a recent QCD calculation  by Frankfurt, Strikman and Zhalov
(FSZ)~\cite{fsz,fs} is motivated by the QCD factorisation theorem for large
$|t|$ rapidity-gap processes and by the correspondence to exclusive
$J/\psi$ production at $|t| \sim 1\gev^2$. In this QCD calculation
in the triple-Pomeron limit, the $W$ dependence of the cross section 
mainly depends on the gluon distribution of the proton.

In all models, a non-relativistic approximation of the $J/\psi$ 
wave-function assuming equal sharing of longitudinal momenta between the
quark and the anti-quark was used. The $J/\psi$ retains the helicity
of the photon which means that $s$-channel helicity is conserved
(SCHC).

The DGLAP-motivated calculation predicts a mild $W$ dependence of the
cross section in the region of  small $|t|$. 
In the region of larger $|t|$ the hard scale is chosen such that saturation is reached 
and thus the cross section is independent of $W$.
The BFKL LL calculations predict a fast rise of the cross section with
$W$ which hardly depends on $|t|$. This is a unique feature of  BFKL
dynamics. The nonL BFKL model behaves in a similar way.
In case of the FSZ parameterisation, the main energy dependence is
provided by the behaviour of the gluon distribution.

All calculations predict an approximate power-law $t$-dependence of
the cross section of the form $d\sigma/dt \sim |t|^{-n}$, where the value of $n$ may depend on 
the  $|t|$ range.

\section {Monte Carlo and background evaluation}

The acceptance and the effects of the detector response were
determined using Monte Carlo (MC) events. All generated events were
passed through the standard ZEUS detector simulation, based on \textsc{Geant 3.13}~\cite{geant}, the ZEUS trigger-simulation
package and  the same reconstruction and analysis  programs as used for the data.

\subsection{The process \boldmath$e p\rightarrow e~J/\psi~Y$}

The process $e p\rightarrow e~J/\psi~Y$ was modelled using the
\textsc{Epsoft} generator~\cite{kasprzak, adamczyk}.  The $\gamma p$
interactions were simulated assuming the exchange of a colourless
object which couples to the whole proton, which subsequently fragments
into a state $Y$. The particle multiplicities and the transverse
momenta of the hadrons in the final state $Y$ were simulated using
parameterisations of $pp$ data, while the longitudinal momenta were
generated with a uniform rapidity distribution.  The
differential cross-section $d\sigma/dt$  was 
reweighted to obtain the shape observed in data. The assumption of $s$-channel helicity conservation 
  was applied.

The differential cross section in $M_Y^2$ have the form $d\sigma/dM_Y^2
\propto (M_Y^2)^{-\beta(t,W)}$. 
The measured $z$ distributions in $|t|$ and $W$ bins were used to determine 
the $|t|$ and $W$ dependence of the function $\beta(t,W)$.
For each bin, a single value of the function $\beta$ was
extracted by using a $\chi^2$ minimisation method. The results were
parameterised in the form of
$\beta(t,W)=(W/W_0)^{0.52\pm0.11}\exp((0.08\pm 0.07) +(-0.14\pm
0.03)|t|)$, with $W_0=95\gev$.  This parameterisation was used in all
further studies.

\subsection{Evaluation of background}

The main sources of background were the non-resonant
QED $\gamma \gamma$ processes, misidentified
pion production and resonant background produced through the decay of
the $\psi(2S)$ meson.

The non-resonant background due to the QED Bethe-Heitler di-muon
production, $ep \rightarrow e\mu^+\mu^-Y$, was simulated using the
\textsc{Grape-Dilepton 1.1} generator \cite{cpc:136:126}.  
The background from $\gamma \gamma \rightarrow \mu^+\mu^- $ events was
estimated in each bin by normalising to the luminosity of the
data. The contribution of this background increased with $|t|$ from
$6$ to $10 \%$.

The $\psi(2S)$ background was estimated using the \textsc{Dipsi}
generator~\cite{cpc:100:195}.  This background was dominated by the
%processes $\psi\rightarrow\pi^0\pi^0J/\psi$ (BR=$(16.84\pm0.33)\%$)
%and $\psi\rightarrow\mu^+\mu^-$ (BR=$(0.75\pm0.08)\%$).  It amounted to about  $1\%$ 
%and $0.1\%$, respectively.
processes $\psi\rightarrow J/\psi\pi^0\pi^0$ (BR=$(17.51\pm0.34)\%$)
and $\psi\rightarrow\mu^+\mu^-$ (BR=$(0.76\pm0.08)\%$).  It amounted to about  $1\%$ 
and $0.1\%$, respectively. 
%new
The contribution from $\psi\rightarrow J/\psi\pi^+\pi^-$ (BR=$(33.1\pm0.5)\%$) was strongly suppressed by selection criteria, i.e. a requirement of two-track events.

The background from exclusive $J/\psi$ production was
found~\cite{epj:c26:389} to be $5\%$ for $2<|t|<3\gev^2$.
 For $|t|>3\gev^2$, it was found to be consistent with zero. All background processes were subtracted bin-by-bin.

\section{Systematic uncertainties}

The systematic uncertainties were determined by
varying the selection cuts and modifying the analysis procedure. Their effects on the integrated cross section are given in parentheses:
\begin{itemize}
\item the cut on $Z$ vertex   was changed by $\pm 10 \cm$ ($+1.8\%$, $-0.5\%$);
\item the $\mu^+\mu^-$ mass window was changed to $2.8-3.4\gev$ ($+0.2\%$);
\item instead of using the MC to subtract the background bin-by-bin, it was 
fitted with a polynomial function and statistically subtracted ($+1.5\%$);
\item the minimum  energy of the CAL energy deposit included for the evaluation 
of the $z$ variable was varied by $\pm 100\mev$ ($+0.3\%$, $-1.4\%$);
\item the strategy of matching energy deposits to the decay tracks was changed. 
Instead of matching every object within $30\cm$ from the track, only
one island within this distance was matched to the track ($-0.7\%$);
\item the uncertainty of the muon acceptance, including the detector, the 
trigger and the reconstruction efficiency, was obtained from a
study~\cite{monica} based on an independent dimuon sample ($\pm6.3\%$);
\item the uncertainty on the acceptance due to modelling of the hadronic final 
state in the \textsc{Epsoft} MC was estimated by varying the parameter
$\beta$ within its errors ($\pm2\%$).
\end{itemize}

The overall systematic uncertainty was determined by adding all the
individual uncertainties in quadrature.  The uncertainty on the
luminosity measurement, $2\%$, was not included.
%--------------------
\section{Results}
%--------------------

\subsection{The \boldmath$J/\psi$ signal}

\begin{figure}[t]
\begin{center}
\includegraphics[scale=.85]{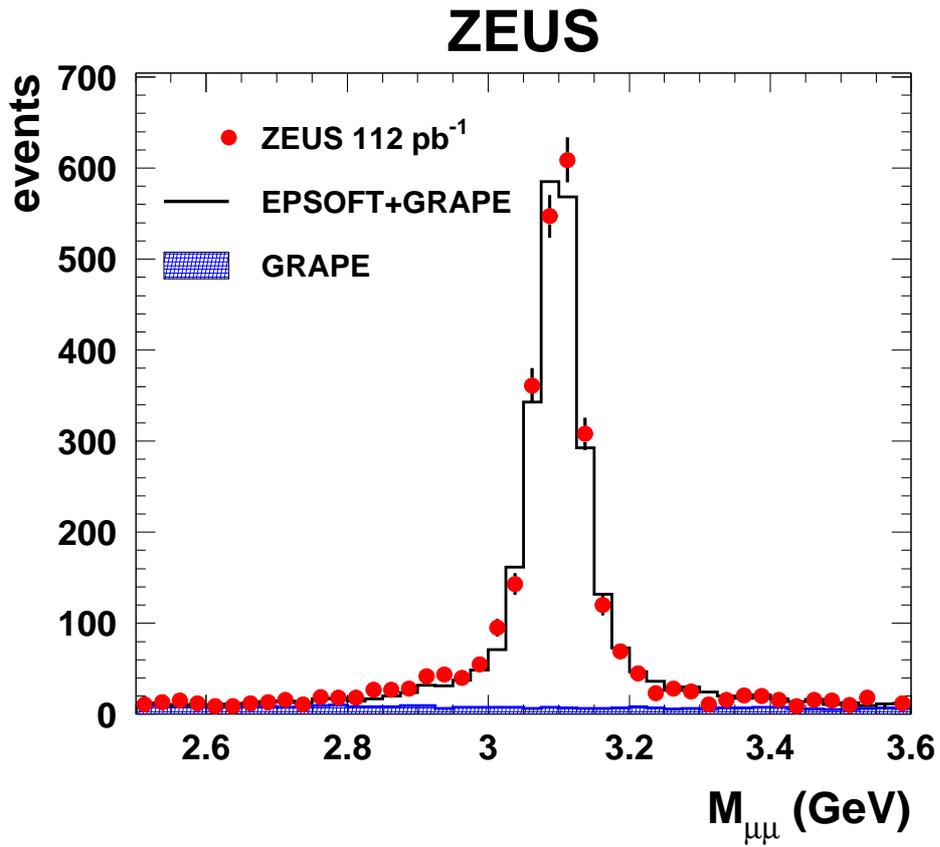}
\end{center}
\caption{
The invariant-mass spectrum for $\mu^+ \mu^-$ pairs in the range
$30<W<160\gev$, $2<|t|<20\gev^2$ and $z>0.95$. Error bars represent only statistical uncertainties.
  The data are compared
to the MC distributions. The hatched histogram represents the $ep
\rightarrow e\mu^+\mu^-Y$ background as simulated by the \textsc{Grape}
MC.  The solid-line histogram represents the sum of $J/\psi$ and background
MC events.
\label{fig-fig3}
}
\end{figure}

The invariant-mass distribution of the $\mu^+\mu^-$ pairs is presented
in \fig{fig3}. A clear peak at the $J/\psi$ mass is observed with very
little non-resonant background.

\begin{figure}[t]
\begin{center}
\includegraphics[scale=.85]{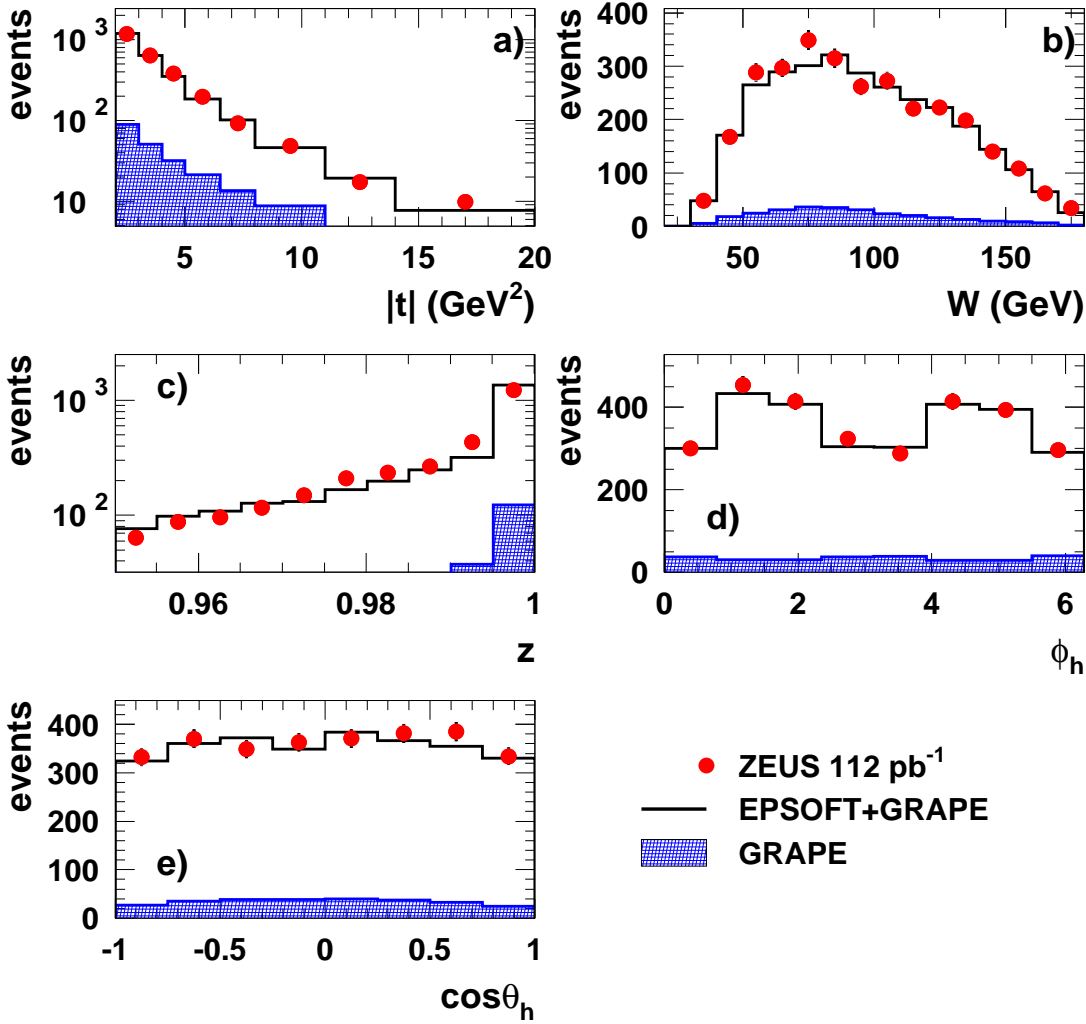}
\end{center}
\caption{
Comparison between the data and MC distributions in the range
$30<W<160\gev$, $2<|t|<20\gev^2$ and $z>0.95$ for a$)$ $|t|$, b$)$
$W$, c$)$ $z$, d$)$ $\phi_h$, e$)$ $\cos \theta_h$.  Error bars represent only statistical uncertainties. 
The hatched histograms represent the $ep
\rightarrow e\mu^+\mu^-Y$ background as simulated by the \textsc{Grape}
MC.  The solid-line histogram represents the sum of $J/\psi$ and
background MC events.
\label{fig-fig4}
}
\end{figure}

The  distributions of the kinematic variables $|t|$, $W$, $z$, $\phi_h$ and
$\cos\theta_h$ are shown in \fig{fig4}.  The MC distributions of the
$J/\psi$ events are shown as well as the QED background.  The overall
agreement between data and MC is good.

\subsection{Determination of photon-proton cross-section}

The $ep$ cross section was determined by subtracting the background
from the data, correcting for the acceptance, using the branching
ratio for the muon channel decay ($5.88\pm0.10\%$) and using the
measured luminosity.

Photon-proton cross sections were extracted from the $ep$ cross sections by using  photon flux factors.  The flux factors~\cite{prep:15:181} generated at the leptonic vertex relate the
$ep$ and the $\gamma p$ cross sections by
\begin{displaymath}
\frac{d^2\sigma^{ep\rightarrow eJ/\psi Y}}{dydQ^2}=\Gamma_T(y,Q^2)\sigma^{\gamma p}(y),
\end{displaymath}
where $\Gamma_T$ is the effective photon flux. The cross sections for different beam energies were averaged using the corresponding luminosities.

\subsection{\boldmath$|t|$ dependence}
\label{sec-tdep}

Differential $\gamma p$ cross section $d\sigma/dt$ for 
proton-dissociative $J/\psi$ photoproduction was measured in the
kinematic region $30<W<160\gev$, $2<|t|<20\gev^2$ and $z>0.95$.

\begin{figure}[h]
\begin{center}
\includegraphics[scale=.85]{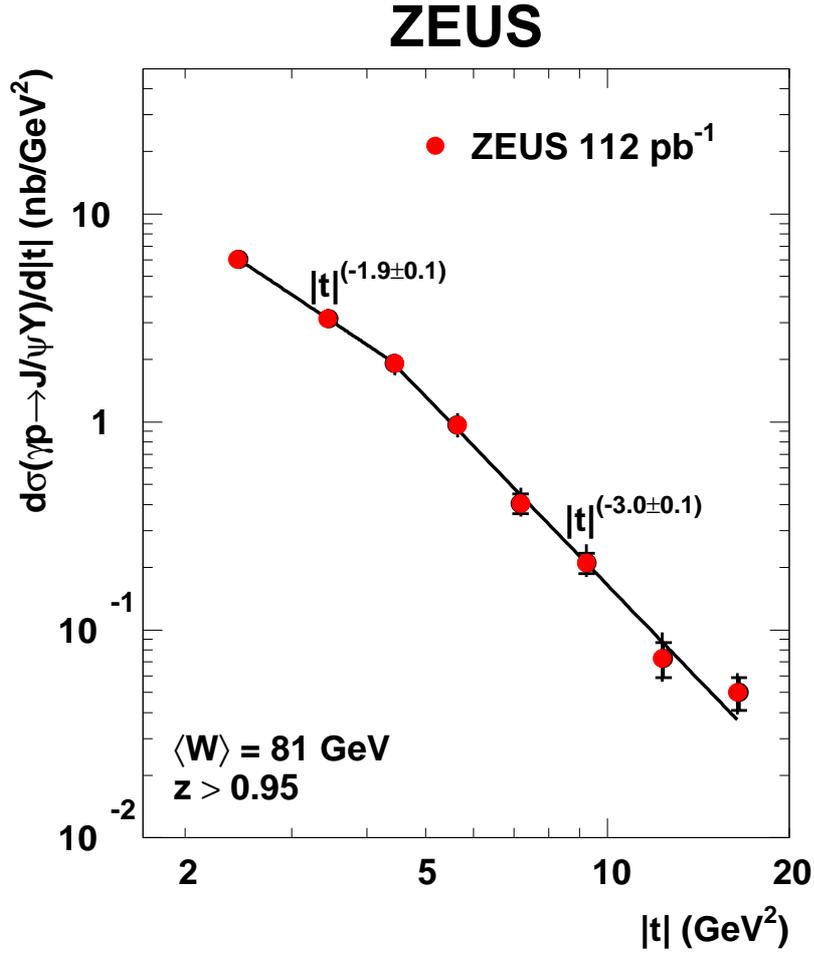}
\end{center}
\caption{
The $|t|$ dependence of the differential cross-section $d\sigma/d|t|$
for the process $\gamma p \rightarrow J/\psi~Y$ at $\langle W\rangle=81\gev$ and
$z>0.95$.  The inner bars correspond to the
statistical uncertainties and the outer to the statistical and
systematic uncertainties added in quadrature. The solid lines are the results of power fits to the form
$d\sigma/dt \sim |t|^n$.
\label{fig-fig5}
}
\end{figure}

The differential cross section as a function of $|t|$ is shown in
\fig{fig5} and listed in \tab{table1}.
The cross section falls steeply with $|t|$.  The data cannot be
described in the whole $|t|$ region by one exponential function of the
form $\sim e^{-b|t|}$, where $b$ is a constant. 
Neither does a single power-law dependence of the form $|t|^{-n}$, where $n$
is a constant,  fit the data. A good fit can, however, be
obtained by fitting two $|t|$ ranges separately, giving
 $n = 1.9 \pm 0.1$ for $ 2<|t|<5\gev^2$ and  $n =
3.0 \pm 0.1$ for $5<|t|<20\gev^2$. 
Note that a good fit can also be 
obtained to a quadratic exponential function $e^{-b|t|+c|t|^2}$.

\begin{figure}[h]
\begin{center}
\includegraphics[scale=.85]{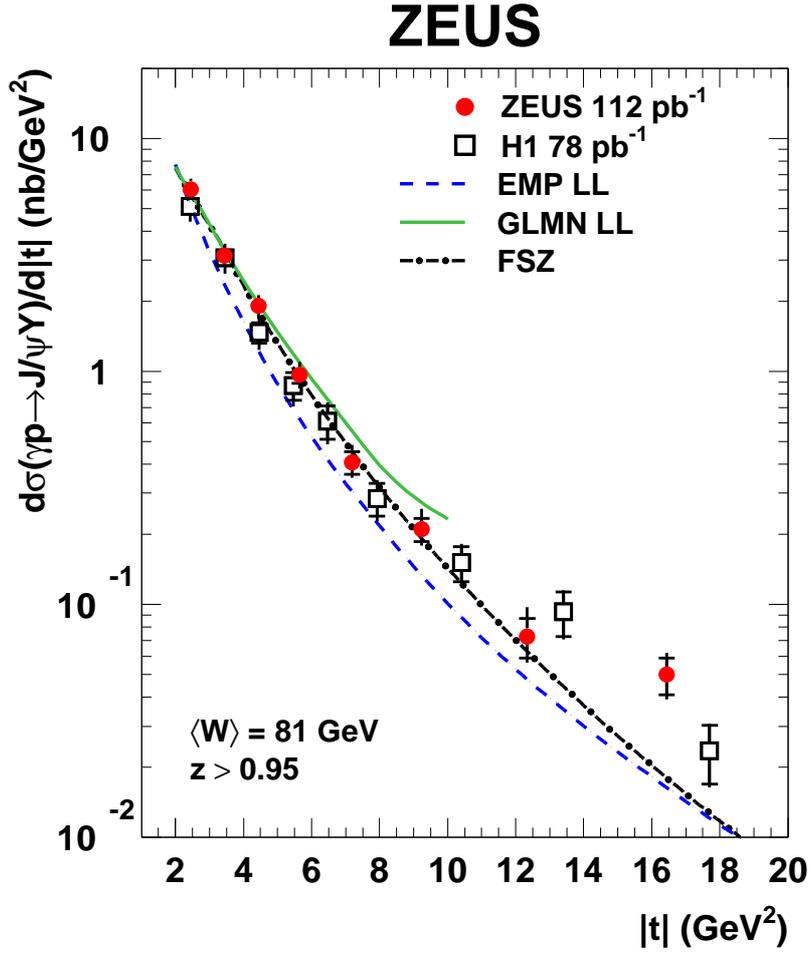}
\end{center}
\caption{
The $|t|$ dependence of the differential cross-section $d\sigma/d|t|$
for the process \ \ \ $\gamma p \rightarrow J/\psi~Y$ \ \
at~$\langle W \rangle=81\gev$ and $z>0.95$. 
The H1 data, $50<W<150\gev$, \protect\cite{H1-03}   are also shown. 
The inner bars correspond to the
statistical uncertainties and the outer to the statistical and
systematic uncertainties added in quadrature. The lines show the predictions of
several calculations, referred to in the text.
\label{fig-fig6}
}
\end{figure}

The differential cross-section  $d\sigma/dt$ as a function of
$|t|$ is shown again in \fig{fig6}, together with the H1 data~\cite{H1-03}, 
and  compared with different
theoretical models.  The GLMN LL model gives a good description of the data
 up to about $|t| = 5\gev^2$, but falls off slower than
the data up to the region where the calculation is valid ($|t|<10\gev ^2$).
  The EMP LL prediction, using $\alpha_S=0.205$ in the
pre-factor and $\alpha_S=0.16$ in the BFKL evolution, lies below the
data in the whole range of $|t|$.  The FSZ results are shown for a
calculation using a Pomeron trajectory with intercept of 1.1 and a
slope of $0.005\gev^{-2}$. Similar results are obtained with a Pomeron
intercept of 1.0.  The CTEQ6M
parameterisation~\cite{cteq} of the parton density functions is
used. The FSZ calculation describes the data well up to $|t|$ of about
$12\gev^2$ but falls-off too steeply at larger $|t|$ values.

\subsection{\boldmath$W$ dependence}

In the Regge formalism, the differential cross section can be
expressed as
\begin{equation}
d\sigma/dt = F(t)W^{4(\apom(t)-1)}, 
\end{equation} 
where $F(t)$ is a function of $t$ and $\apom(t)$ is the effective Pomeron
trajectory.  This expression is usually used for exclusive reactions,
but has been used also for the case where  
 $M_Y$ is integrated over~\cite{epj:c26:389,H1-03}. By studying the $W$
dependence of $d\sigma/dt$ at fixed $t$, the Pomeron trajectory can be determined.

\begin{figure}[h]
\begin{center}
\includegraphics[scale=0.8]{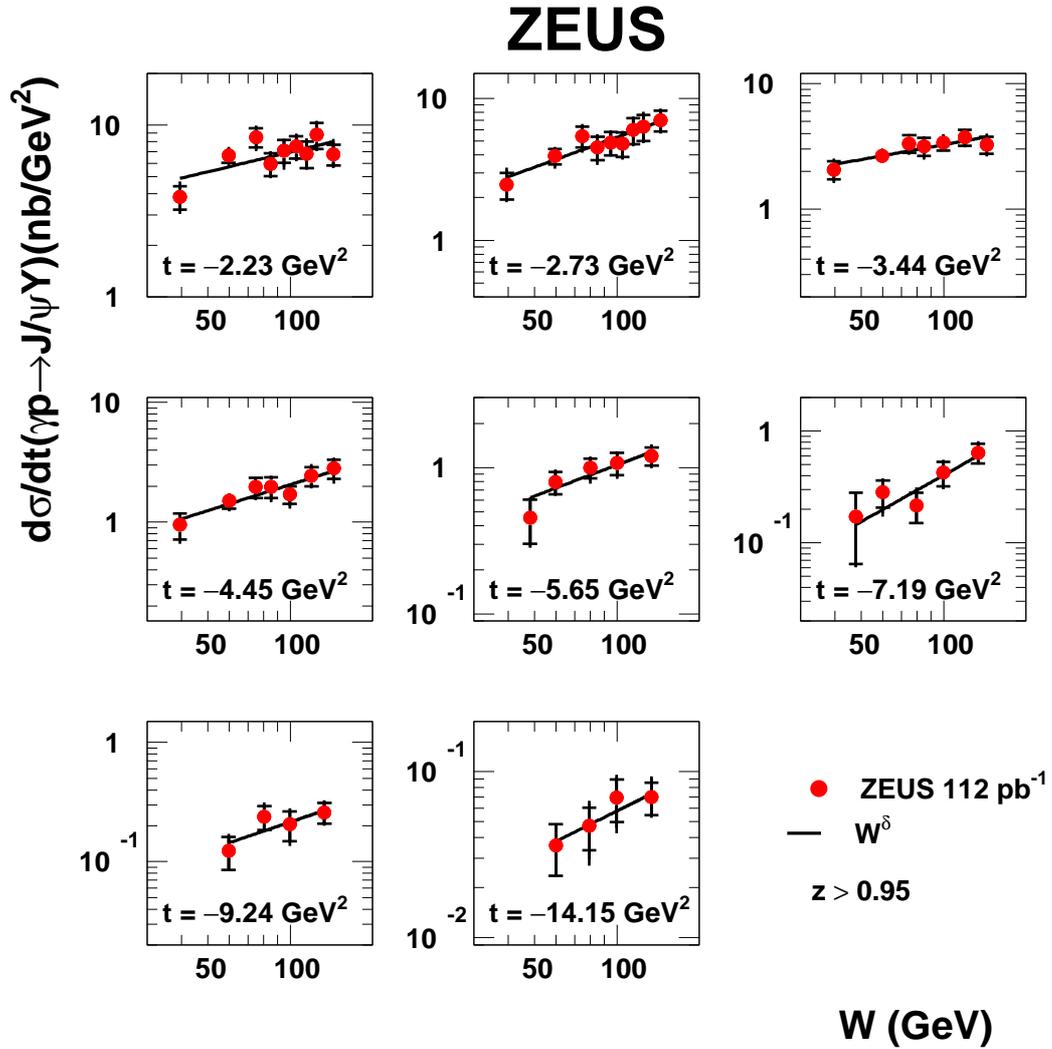}
\end{center}
\caption{
The $W$ dependence of the differential cross-section $d\sigma/dt$ for
the process $\gamma p \rightarrow J/\psi~Y$ ($z>$0.95) at fixed $|t|$
values, as indicated in the figure. The inner bars correspond to the
statistical uncertainties and the outer to the statistical and
systematic uncertainties added in quadrature. The solid lines are the
results of  fits to the form $d\sigma/dt \sim W^\delta$.
\label{fig-fig7}
}
\end{figure}

\begin{figure}[h]
\begin{center}
\includegraphics[scale=0.8]{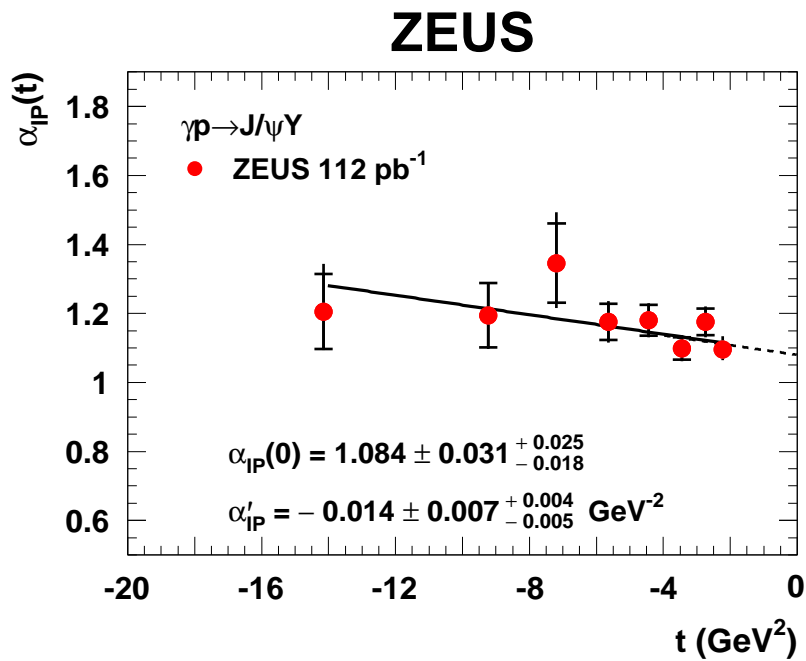}
\end{center}
\caption{
The effective Pomeron trajectory as a function of $t$.  The inner bars
correspond to the statistical uncertainties and the outer to the
statistical and systematic uncertainties added in quadrature. The solid
line is a fit of the form $ \apom(t)=\apom(0)+\aprime\cdot t $. The
dashed line is an extrapolation to $\apom(0)$.
\label{fig-fig8}
}
\end{figure}

The $W$ dependence of the differential cross section $d\sigma/dt$ for
eight fixed $t$ values is shown in \fig{fig7} and listed in
\tab{table2}. At each $t$ value the cross section is parameterised as 
$\sigma \sim W^\delta$ and the lines in the figure are the result of
these fits.
The values of $\apom$ can be obtained at each $t$ value through
\begin{equation}
\apom =(\delta+4)/4 
\end{equation}
and are shown in \fig{fig8}. The values of $\delta$ and of $\apom$
are listed in \tab{table3}.  A linear fit of the form
\begin{equation}
\apom(t)=\apom(0)+\aprime \cdot t 
\end{equation}
yields an intercept 
\begin{equation}
\apom(0)=1.084
\pm{0.031}(\rm{stat.})^{+0.025}_{-0.018}(\rm{syst.}),
\end{equation}
and a slope
\begin{equation}
\aprime=-0.014
\pm{0.007}(\rm{stat.})^{+0.004}_{-0.005}(\rm{syst.}).
\end{equation}
The value of the intercept is consistent with that of
the so-called ``soft" Pomeron~\cite{dl} (1.0808). The slope is
different from that of the ``soft" Pomeron~\cite{dlaprime} ($0.25\gev^{-2}$), 
but is consistent with the predictions of the BFKL
Pomeron\cite{brodsky, nikolaev}.

\begin{figure}[h]
\begin{center}
\includegraphics[scale=0.8]{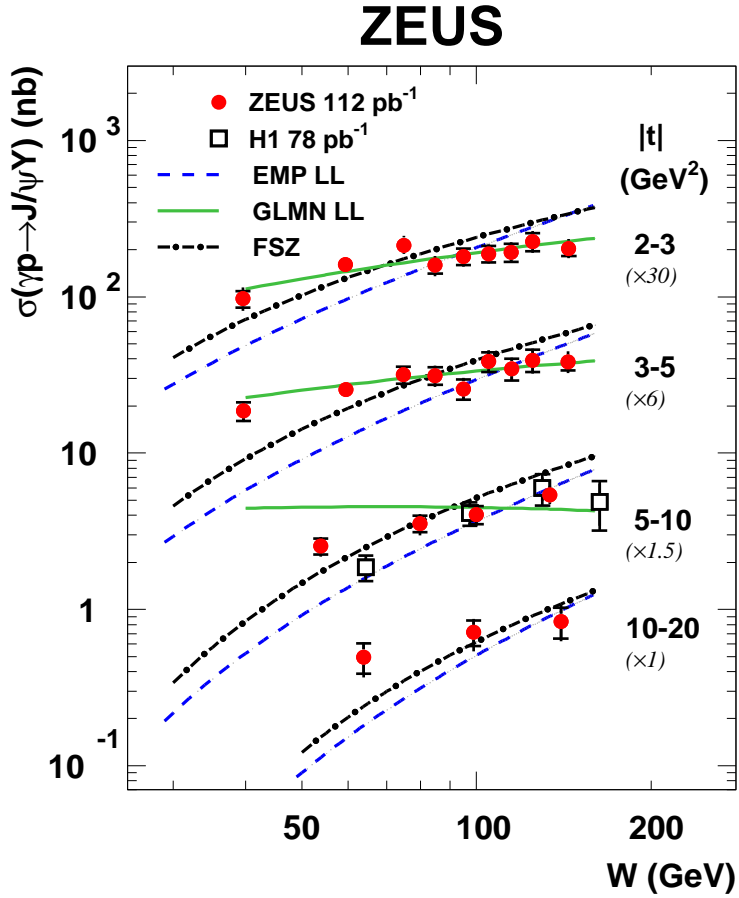}
\end{center}
\caption{
The $W$ dependence for the process $\gamma p \rightarrow J/\psi~Y$
($z>$0.95) in four different $|t|$ bins. 
The H1 data\protect\cite{H1-03} for the $|t|$ bin of $5$ to $10\gev^2$  are also shown. The inner bars
correspond to the statistical uncertainties and the outer to the
statistical and systematic uncertainties added in quadrature. The lines show the
predictions of several calculations referred to in the text.
\label{fig-fig9}
}
\end{figure}

The $\gamma p$ cross section as a function of $W$ was measured in four
bins of $|t|$: $2<|t|<3\gev^2$; $3<|t|<5\gev^2$; $5<|t|<10\gev^2$ and
$10<|t|<20\gev^2$ for $30<W<160\gev$. The cross-section values are
shown in \fig{fig9} and summarised in
\tab{table4}. The H1 data~\cite{H1-03} for the $|t|$
bin of 5 to $10\gev^2$ are also shown. A clear rise with $W$ is seen in all the four $|t|$
regions.  Also shown in the figure are the predictions of the models
 used in the comparison with $d\sigma/dt$. The DGLAP-based GLMN
LL calculation agrees well with the data in the first two $|t|$ bins,
but fails to describe the rise with $W$ for $|t| > 5\gev^2$. The
other two calculations, EMP LL and FSZ, predict a $W$ dependence which
is too steep in all the $|t|$ ranges presented in the analysis.

\subsection{Decay angular distributions}

The angular distributions of the $J/\psi$ decay provide information
about the photon and $J/\psi$ polarisation states.  The normalised
two-dimensional angular distributions can be written in terms of spin
density matrix elements, $r$, as:
\begin{eqnarray}
\lefteqn{ \frac{1}{\sigma}\frac{d^2\sigma}{d\cos\theta_h d\phi_h}=\frac{3}{4\pi}\left ( \frac{1}{2}(1+r^{04}_{00})-\frac{1}{2}(3r^{04}_{00}-1)\cos^2\theta_h+ \right .} & & \nonumber\\
& & \left
. +\sqrt{2}\mathrm{Re}\{r^{04}_{10}\}\sin2\theta_h\cos\phi_h+
r^{04}_{1-1}\sin^2\theta_h\cos2\phi_h\right ).
\label{eq-2dhel}
\end{eqnarray}
 
The one-dimensional distributions result from the integration over
$\theta_h$ or $\phi_h$ and are expressed as :
\begin{equation}
\frac{d\sigma}{d\cos\theta_h}\propto 1+r^{04}_{00}+(1-3r^{04}_{00})\cos^2\theta_h
\label{eq-cos}
\end{equation}
and
\begin{equation}
\frac{d\sigma}{d\phi} \propto 1+ r^{04}_{1-1}\cos 2\phi_h.
\label{eq-phi}
\end{equation}
The spin density matrix element $r^{04}_{00}$ represents the
probability that the produced $J/\psi$ has helicity zero,
$\mathrm{Re}\{r^{04}_{10}\}$ is proportional to the single-flip
amplitude and $r^{04}_{1-1}$ is related to the interference between
non-flip and double-flip amplitudes. If SCHC holds, the $J/\psi$
retains the helicity of the almost real photon and all the three
matrix elements are expected to be zero.

\begin{figure}[h]
\begin{center}
\includegraphics[scale=0.8]{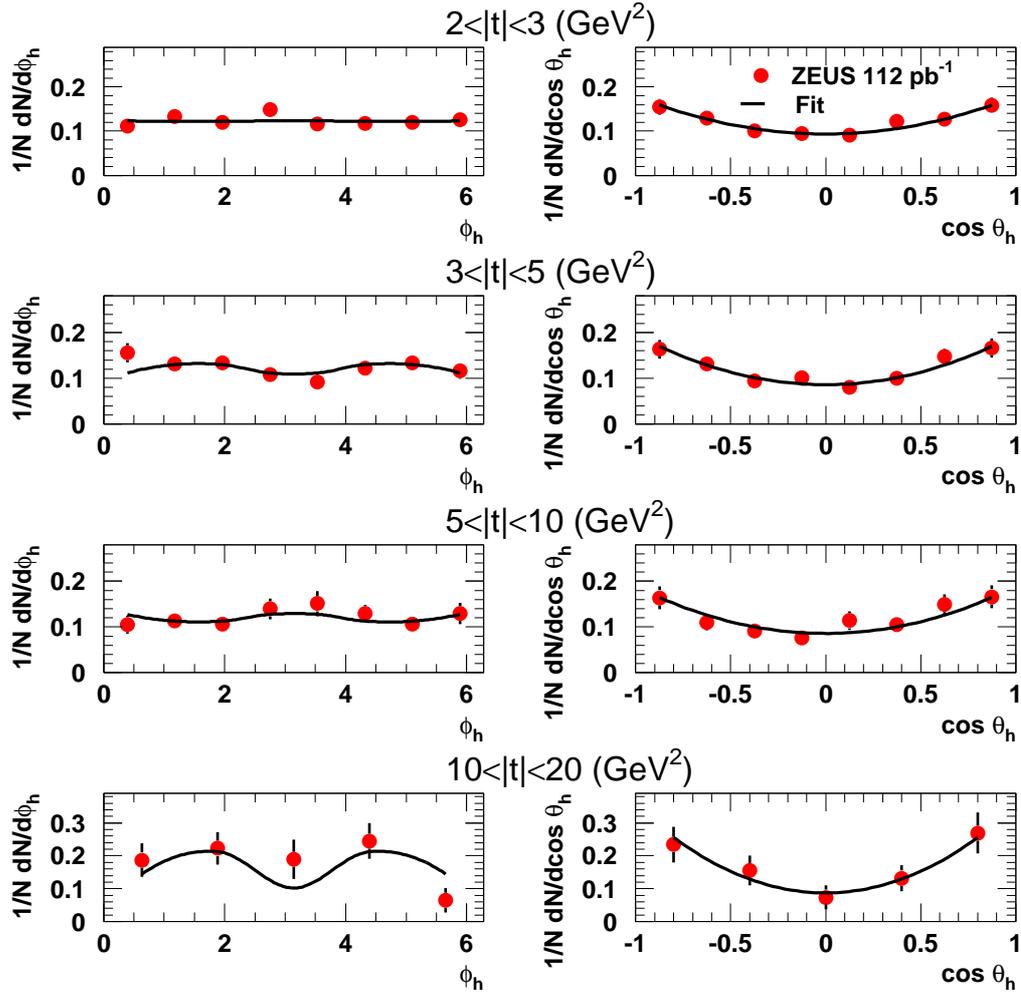}
\end{center}
\caption{The normalized distributions of $\phi$ and $\cos(\theta)$ for $30<W<160\gev$ and $z>0.95$ in 
four bins of $|t|$. Error bars represent only statistical uncertainties. The lines represent the results of the fits according to formulae \eq{cos} and \eq{phi} in the text.
\label{fig-fig10}
}
\end{figure}

\begin{figure}[h]
\begin{center}
\includegraphics[scale=0.8]{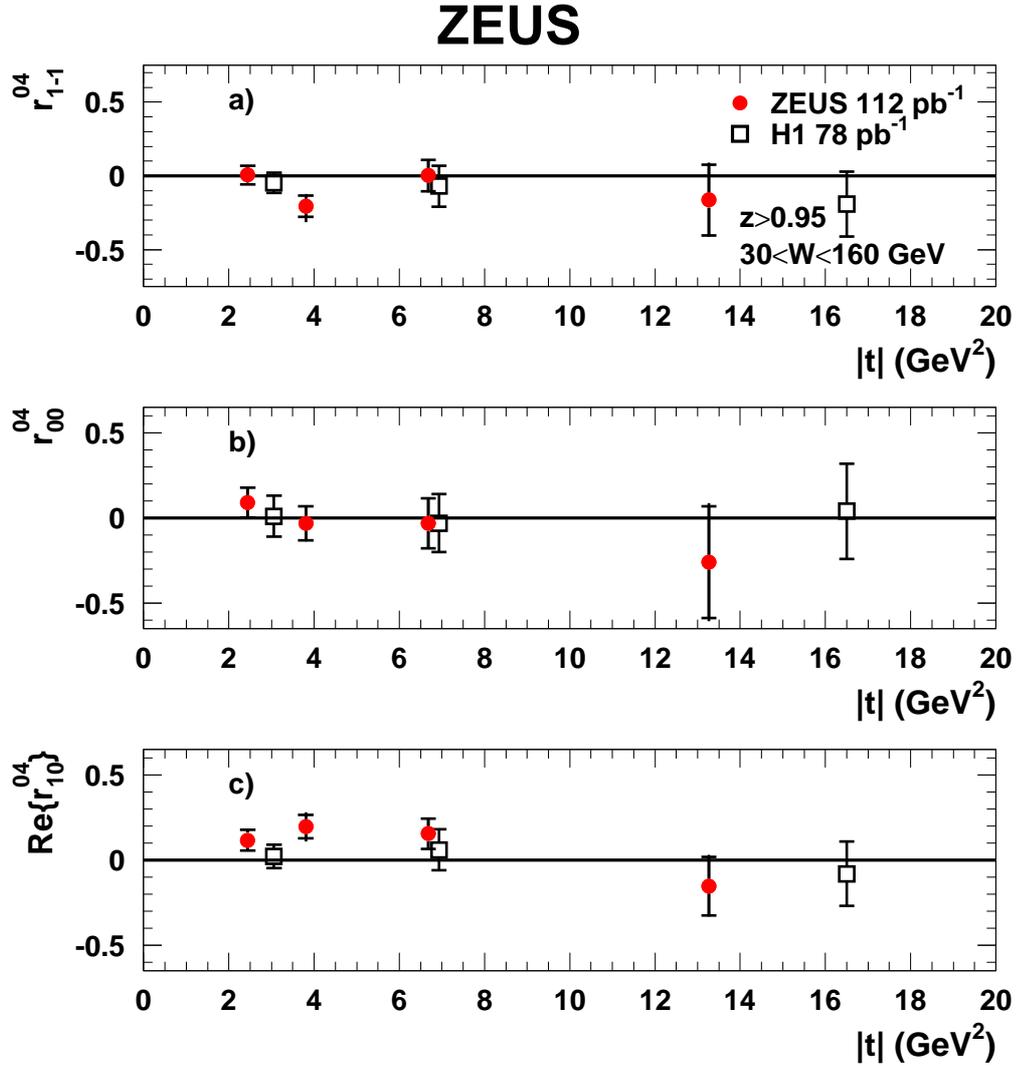}
\end{center}
\caption{Helicity spin density matrix elements 
a) $r^{04}_{1-1}$, b) $r^{04}_{00}$ and c)
$\mathrm{Re}\{r^{04}_{10}\}$ as a function of $|t|$ in the range
$30<W<160\gev$ and $z>$0.95. 
The H1 data, $50<W<150\gev$, \protect\cite{H1-03}  are also shown. 
The inner bars correspond to the
statistical uncertainties and the outer to the statistical and
systematic uncertainties added in quadrature. The solid lines show the
expectation from SCHC.
\label{fig-fig11}
}
\end{figure}

The distributions of $\cos\theta_h$ and $\phi_h$ after background
subtraction and acceptance corrections  
 in four $|t|$ bins and for $30<W<160\gev$ are shown in \fig{fig10}.  
They were fitted using
formulae \eq{cos} and \eq{phi}.  The $r^{04}_{00}$,
$\mathrm{Re}\{r^{04}_{10}\}$ and $r^{04}_{1-1}$ spin density matrix
elements were extracted from a two-dimensional $\chi^2$ minimisation fit using Eq.~\eq{2dhel} 
and are summarised in \tab{table5} and shown in
\fig{fig11}. The H1 data~\cite{H1-03}
are also shown. The results for $r^{04}_{00}$ and $r^{04}_{1-1}$ are compatible with zero. 
The values obtained for $\mathrm{Re}\{r^{04}_{10}\}$ are not compatible with 
zero for $|t|<10\gev^2$, contrary to the expectation from SCHC.

The measurements of the present analysis,  the differential 
cross section as a function of $|t|$, the $W$ dependence of the cross
section and the density matrix elements, are in good agreement with those of the H1
collaboration~\cite{H1-03} within the common kinematic region. 
%---------------------------------
%-----------------
\section{Summary}
%------------------

Proton-dissociative $J/\psi$ production was measured at HERA in the
photoproduction regime in the kinematic region $30<W<160\gev$,
$z>0.95$ and $2<|t|<20\gev^2$.

The $|t|$ dependence of the differential cross section, $d\sigma/d|t|$, is
found to be approximately power-like, $\sim |t|^{-n}$, with the power
$n$ increasing with $|t|$.

The effective Pomeron trajectory was derived from a measurement of the
$W$ dependence of the cross section at fixed $t$ values. The value
of the slope of the trajectory is compatible with zero. It is
consistent with the predictions of the BFKL Pomeron but different from
the slope of the ``soft" Pomeron.

The cross-section $\sigma(\gamma p \to J/\psi \ Y)$ rises significantly
with $W$ in each $|t|$ bin.
The $t$ and $W$ dependence of the cross section were compared to several
theoretical calculations. The DGLAP-motivated GLMN LL~\cite{glmn} calculation can
describe the behaviour of the data, both in $t$ and in $W$, up to
$|t| = 5\gev^2$. The BFKL-motivated EMP LL~\cite{enberg} calculation fails to
describe the data in the kinematic region of the present
measurement. The FSZ~\cite{fsz,fs} calculation describes the $t$ dependence of the
cross section only  up to $|t| = 12\gev^2$ and fails to reproduce the $W$
dependence.

The spin density matrix elements of the $J/\psi$, $r^{04}_{00}$ 
and $r^{04}_{1-1}$ are consistent with
zero, as expected from $s$-channel helicity conservation.
The values obtained for $\mathrm{Re}\{r^{04}_{10}\}$ are not compatible with
zero for $|t|<10\gev^2$, contrary to the expectation from SCHC. 
\section*{Acknowledgments}

We appreciate the contributions to the construction and maintenance of the ZEUS
detector of many people who are not listed as authors. 
The HERA machine group and the DESY computing staff are especially 
acknowledged for their success in providing excellent operation of 
the collider and the data analysis environment. We thank the DESY 
directorate for their strong support and encouragement.
We want to thank Asher Gotsman, Jeff Forshaw, Lonya
Frankfurt, Uri Maor, Leszek Motyka and Mark Strikman for many useful
discussions. We are grateful to Rikard Enberg, Eran Naftali and
Michael Zhalov for providing the results of their theoretical
calculations.
\vfill\eject

%------------------------------------------------------------------------------
%       Bibliography
%------------------------------------------------------------------------------
{
\def\bibname{\Large\bf References}
\def\refname{\Large\bf References}
\pagestyle{plain}
\ifzeusbst
  \bibliographystyle{./BiBTeX/bst/l4z_default}
\fi
\ifzdrftbst
  \bibliographystyle{./BiBTeX/bst/l4z_draft}
\fi
\ifzbstepj
  \bibliographystyle{./BiBTeX/bst/l4z_epj}
\fi
\ifzbstnp
  \bibliographystyle{./BiBTeX/bst/l4z_np}
\fi
\ifzbstpl
  \bibliographystyle{./BiBTeX/bst/l4z_pl}
\fi
{\raggedright
\bibliography{./BiBTeX/user/syn.bib,%
              ./BiBTeX/bib/l4z_articles.bib,%
              ./BiBTeX/bib/l4z_books.bib,%
              ./BiBTeX/bib/l4z_conferences.bib,%
              ./BiBTeX/bib/l4z_h1.bib,%
              ./BiBTeX/bib/l4z_misc.bib,%
              ./BiBTeX/bib/l4z_old.bib,%
              ./BiBTeX/bib/l4z_preprints.bib,%
              ./BiBTeX/bib/l4z_replaced.bib,%
              ./BiBTeX/bib/l4z_temporary.bib,%
              ./BiBTeX/bib/l4z_zeus.bib,%
              ./BiBTeX/user/user.bib}}
}
\vfill\eject

%------------------------------------------------------------------------------
%       Tables
%------------------------------------------------------------------------------
%-------------------------------------------------------------------------------
%       An example table
%-------------------------------------------------------------------------------
\begin{table}[p]
\begin{center}
\begin{tabular}{|rp{.1cm}r|p{.3cm}rp{.3cm}|c|}
\hline
 \multicolumn{3}{|c|}{$|t|$ bin}& 
  \multicolumn{3}{c|}{$\langle|t|\rangle$}&
 $d\sigma/d|t|$ \\
 \multicolumn{3}{|c|}{ ($\gev^2$)}& 
  \multicolumn{3}{c|}{  ($\gev^2$)}&
  $(\nb/\gev^2)$ \\
\hline
\hline
2.0&$-$&3.0   & &2.5 &  &6.05 $\pm$ 0.23 $^{+0.60}_{-0.43}$   \\
3.0&$-$&4.0   & &3.5 &  &3.14 $\pm$ 0.18 $^{+0.29}_{-0.23}$    \\
4.0&$-$&5.0   & &4.4 &  &1.91 $\pm$ 0.13 $^{+0.17}_{-0.18}$     \\
5.0&$-$&6.5   & &5.6 &  &0.97 $\pm$ 0.08 $^{+0.11}_{-0.10}$     \\
6.5&$-$&8.0   & &7.2 &  &0.41 $\pm$ 0.05 $^{+0.06}_{-0.03}$     \\
8.0&$-$&11.0  & &9.3 &  &0.21 $\pm$ 0.02 $^{+0.04}_{-0.02}$    \\
11.0&$-$&14.0 & &12.4&  &0.07 $\pm$ 0.01 $^{+0.02}_{-0.01}$    \\
14.0&$-$&20.0 & &16.5&  &0.05 $\pm$ 0.01 $^{+0.01}_{-0.01}$  \\
\hline

\end{tabular}
\caption{Differential cross-section  $d\sigma/d|t|$ as a function of $|t|$
for $30<W<160\gev$ and $z>0.95$. The first uncertainty is statistical
and the second is systematic.}
\label{tab-table1}
\end{center}
\end{table}
%-------------------------
\begin{table}[p]
\begin{center}
\begin{tabular}{|c|c|rb{1pt}r|r@{.}l|c|}
\hline
$|t|$ bin &$-t$  & \multicolumn{3}{c|}{$W$ bin}   &  \multicolumn{2}{c|}{$\langle W\rangle$}   & $d\sigma/dt$ \\
($\gev^2$)& ($\gev^2$) & \multicolumn{3}{c|}{($\gev$)}   & \multicolumn{2}{c|}{($\gev$)} &  $(\nb/\gev^2)$\\

\hline
\hline
\multirow{9}{*}{2.0$-$2.5} & \multirow{9}{*}{2.2} & 30 & $-$ & 50  & 39&5   & 3.82  $\pm$ 0.59  $^{+0.68 }_{-0.51 }$ \\
                                              &    & 50 & $-$ & 70  & 59&5   & 6.65  $\pm$ 0.61  $^{+0.58 }_{-0.65 }$\\
                                              &    & 70 & $-$ & 80  & 75&0   & 8.52  $\pm$ 1.10  $^{+0.90 }_{-0.98 }$ \\
                                              &    & 80 & $-$ & 90  & 84&8   & 5.95  $\pm$ 0.91  $^{+0.65 }_{-0.51 }$ \\
                                              &    & 90 & $-$ & 100 & 95&0   & 7.10  $\pm$ 1.08  $^{+0.57 }_{-1.18 }$ \\
                                              &    &100 & $-$ & 110 & 105&1  & 7.51  $\pm$ 1.13  $^{+0.89 }_{-0.70 }$ \\
                                              &    &110 & $-$ & 120 & 114&9  & 6.83  $\pm$ 1.20  $^{+0.99 }_{-0.58 }$ \\
                                              &    &120 & $-$ & 130 & 125&1  & 8.79  $\pm$ 1.52  $^{+0.98 }_{-0.78}$ \\
                                              &    &130 & $-$ & 160 & 144&0  & 6.74  $\pm$ 0.93  $^{+1.25 }_{-0.63 }$ \\
\hline
\multirow{9}{*}{2.5$-$3.0} & \multirow{9}{*}{2.7 } & 30 & $-$ & 50  & 39&6   & 2.47  $\pm$ 0.52  $^{+0.60 }_{-0.31 }$ \\
                                                &  & 50 & $-$ & 70  & 59&3   & 3.93  $\pm$ 0.50  $^{+0.37 }_{-0.53 }$\\
                                                &  & 70 & $-$ & 80  & 74&8   & 5.43  $\pm$ 0.91  $^{+0.74 }_{-0.91 }$ \\
                                                &  & 80 & $-$ & 90  & 85&0   & 4.52  $\pm$ 0.84  $^{+0.67 }_{-0.73 }$ \\
                                                &  & 90 & $-$ & 100 & 95&0   & 4.89  $\pm$ 0.93  $^{+0.56 }_{-0.32 }$ \\
                                                &  &100 & $-$ & 110 & 104&8  & 4.81  $\pm$ 0.95  $^{+0.51 }_{-0.53 }$ \\
                                                &  &110 & $-$ & 120 & 114&7  & 6.02  $\pm$ 1.26  $^{+0.71 }_{-0.64 }$ \\
                                                &  &120 & $-$ & 130 & 124&9  & 6.34  $\pm$ 1.33  $^{+1.08 }_{-0.65 }$ \\
                                                &  &130 & $-$ & 160 & 144&5  & 7.00  $\pm$ 1.17  $^{+0.92 }_{-0.68 }$ \\
\hline
\multirow{7}{*}{3.0$-$4.0} & \multirow{7}{*}{3.4 } & 30 & $-$ & 50 & 39&7  & 2.06    $\pm$ 0.34   $^{+0.29 }_{-0.30 }$ \\
                                                &  & 50 & $-$ & 70 & 59&5  & 2.65    $\pm$ 0.30  $^{+0.23 }_{-0.13 }$\\
                                                &  & 70 & $-$ & 80 & 74&8  & 3.33    $\pm$ 0.53  $^{+0.30 }_{-0.33 }$ \\
                                                &  & 80 & $-$ & 90 & 84&8  & 3.17    $\pm$ 0.53  $^{+0.34 }_{-0.45 }$ \\
                                                &  & 90 & $-$ & 110& 99&8  & 3.38    $\pm$ 0.45  $^{+0.36 }_{-0.21 }$ \\
                                                &  &110 & $-$ & 130& 119&6 & 3.72    $\pm$ 0.54  $^{+0.32 }_{-0.30 }$ \\
                                                &  &130 & $-$ & 160& 143&5 & 3.26    $\pm$ 0.52  $^{+0.49 }_{-0.38 }$ \\
\hline
\end{tabular}
\caption{Differential cross section as a function of $W$ in eight $t$ bins and for $z>0.95$. The first uncertainty is  statistical and the second is systematic.}
\label{tab-table2}
\end{center}
\end{table}

\begin{table}[p]
\begin{center}
\begin{tabular}{|c|c|rb{1pt}r|r@{.}l|c|}
\hline
$|t|$ bin &$-t$  & \multicolumn{3}{c|}{$W$ bin}   &  \multicolumn{2}{c|}{$\langle W\rangle$}   & $d\sigma/dt$ \\
($\gev^2$)& ($\gev^2$) & \multicolumn{3}{c|}{($\gev$)}   & \multicolumn{2}{c|}{($\gev$)} &  $(\nb/\gev^2)$\\
\hline
\hline
\multirow{7}{*}{4.0$-$5.0} & \multirow{7}{*}{4.5 } & 30 & $-$ & 50  & 39&5 & 0.95   $\pm$ 0.23     $^{+0.08 }_{-0.18 }$ \\
                                                &  & 50 & $-$ & 70  & 59&8 & 1.51   $\pm$ 0.22     $^{+0.10 }_{-0.17 }$\\
                                                &  & 70 & $-$ & 80  & 74&8 & 1.97   $\pm$ 0.38     $^{+0.15 }_{-0.22 }$ \\
                                                &  & 80 & $-$ & 90  & 85&1  & 1.98   $\pm$ 0.39     $^{+0.23 }_{-0.34 }$ \\
                                                &  & 90 & $-$ & 110 & 99&6 & 1.71   $\pm$ 0.29     $^{+0.24 }_{-0.30 }$ \\
                                                &  &110 & $-$ & 130 & 119&7 & 2.44   $\pm$ 0.44    $^{+0.44 }_{-0.29 }$ \\
                                                &  &130 & $-$ & 160 & 144&5 & 2.82   $\pm$ 0.51    $^{+0.37 }_{-0.52 }$ \\
\hline
\multirow{5}{*}{5.0$-$6.5} & \multirow{5}{*}{5.7 } & 30 & $-$ & 70  &  48&1  & 0.45    $\pm$ 0.15  $^{+0.08 }_{-0.07 }$ \\
                                                && 50 & $-$ & 70  &  59&6  & 0.79   $\pm$  0.14  $^{+0.11 }_{-0.07 }$ \\
                                                && 70 & $-$ & 90  &  79&9  & 1.00   $\pm$  0.15  $^{+0.12 }_{-0.16 }$ \\
                                                && 90 & $-$ & 110 & 100&0  & 1.07   $\pm$  0.19  $^{+0.12 }_{-0.15 }$ \\
                                                &&110 & $-$ & 160 &  134&1 & 1.20  $\pm$  0.17  $^{+0.17 }_{-0.13 }$ \\
\hline
\multirow{5}{*}{6.5$-$8.0} & \multirow{5}{*}{7.2 } & 30 & $-$ & 70  &  47&7  & 0.17    $\pm$ 0.11  $^{+0.02 }_{-0.02 }$ \\
                                                && 50 & $-$ & 70  &  60&0  & 0.28   $\pm$  0.08  $^{+0.05 }_{-0.08 }$ \\
                                                && 70 & $-$ & 90  &  79&4  & 0.22   $\pm$  0.07  $^{+0.07 }_{-0.01 }$ \\
                                                && 90 & $-$ & 110 &  99&8  & 0.43   $\pm$  0.11  $^{+0.08 }_{-0.06 }$ \\
                                                &&110 & $-$ & 160 &  133&6 & 0.64  $\pm$  0.13  $^{+0.12 }_{-0.06 }$ \\
\hline
\multirow{4}{*}{8.0$-$11.0} & \multirow{4}{*}{9.2 } & 50 & $-$ & 70  &  59&5  & 0.12    $\pm$ 0.04  $^{+0.03 }_{-0.01 }$ \\
                                                && 70 & $-$ & 90 &  80&4  & 0.24   $\pm$  0.05  $^{+0.05 }_{-0.03 }$ \\
                                                && 90 & $-$ & 110 &  99&6  & 0.21   $\pm$  0.06  $^{+0.05 }_{-0.04 }$ \\
                                                &&110 & $-$ & 160 &  133&4 & 0.26  $\pm$  0.05  $^{+0.04 }_{-0.03 }$ \\
\hline
\multirow{4}{*}{11.0$-$20.0} & \multirow{4}{*}{14.2 } & 50 & $-$ & 70  &  59&8  & 0.04    $\pm$ 0.01  $^{+0.01 }_{-0.01 }$ \\
                                                && 70 & $-$ & 90 &  79&3  & 0.05   $\pm$  0.01  $^{+0.01 }_{-0.01 }$ \\
                                                && 90 & $-$ & 110 & 100&0  & 0.07   $\pm$  0.02  $^{+0.02 }_{-0.02 }$ \\
                                                &&110 & $-$ & 160 &  133&9 & 0.07  $\pm$  0.02  $^{+0.02 }_{-0.01 }$ \\
\hline
\end{tabular}
%\caption[Table 2]{Continuation: Cross section as a function of $W$ in eight $t$ bins and for $z>0.95$. The first uncertainty is  statistical and the second is systematic.}
\end{center}
\vspace{.3cm}
{\bf Table 2 (Continuation):} {\it Differential cross section as a function of $W$ in eight $t$ bins and for $z>0.95$. The first uncertainty is  statistical and the second is systematic.}
\label{table2b}
\end{table}

%-------------------------
\begin{table}[p]
\begin{center}
\begin{tabular}{|rp{1pt}r|cr@{.}p{.7cm}|c|c|c|}
\hline
\multicolumn{3}{|c|}{$|t|$ bin}  &\multicolumn{3}{c|}{$-t$}  & \multirow{2}{*}{$\delta$}   & \multirow{2}{*}{$\apom$}   \\
 \multicolumn{3}{|c|}{($\gev^2$)} & \multicolumn{3}{c|}{($\gev^2$)} & &   \\
\hline
\hline
 2.0&$-$&2.5  && 2&2  & 0.38  $\pm$ 0.10   $^{+0.11 }_{-0.06 }$ & 1.10   $\pm$ 0.03   $^{+0.03 }_{-0.02 }$ \\
 2.5&$-$&3.0  && 2&7  & 0.70  $\pm$ 0.15   $^{+0.10 }_{-0.10 }$ & 1.18   $\pm$ 0.04   $^{+0.02 }_{-0.02 }$ \\
 3.0&$-$&4.0  && 3&4  & 0.39  $\pm$ 0.13   $^{+0.08 }_{-0.07 }$ & 1.10   $\pm$ 0.03   $^{+0.02 }_{-0.02 }$ \\
 4.0&$-$&5.0  && 4&5  & 0.72  $\pm$ 0.18   $^{+0.08 }_{-0.07 }$ & 1.18   $\pm$ 0.05   $^{+0.02 }_{-0.02 }$ \\
 5.0&$-$&6.5  && 5&7  & 0.70  $\pm$ 0.21   $^{+0.12 }_{-0.12 }$ & 1.18   $\pm$ 0.05   $^{+0.03 }_{-0.03 }$ \\
 6.5&$-$&8.0  && 7&2  & 1.38  $\pm$ 0.46   $^{+0.37 }_{-0.24 }$ & 1.35   $\pm$ 0.12   $^{+0.09 }_{-0.06 }$ \\
 8.0&$-$&11.0 && 9&2  & 0.78  $\pm$ 0.38   $^{+0.06 }_{-0.12 }$ & 1.20   $\pm$ 0.09   $^{+0.02 }_{-0.03 }$ \\
11.0&$-$&20.0 && 14&2 & 0.82  $\pm$ 0.44   $^{+0.34 }_{-0.08 }$ & 1.21   $\pm$ 0.11   $^{+0.09 }_{-0.02 }$ \\
\hline
\end{tabular}
\caption{The values of the parameters $\delta$ and the effective Pomeron trajectory
$\apom$ for eight fixed $t$ values.  The first uncertainty is statistical
and the second is systematic.}
\label{tab-table3}
\end{center}
\end{table}

%-------------------------
\begin{table}[p]
\begin{center}
\begin{tabular}{|c|c|rb{1pt}r|r@{.}l|c|}
\hline
$|t|$ bin &$-t$  & \multicolumn{3}{c|}{$W$ bin}   &  \multicolumn{2}{c|}{$\langle W\rangle$}   & $d\sigma/dt$ \\
($\gev^2$)& ($\gev^2$) & \multicolumn{3}{c|}{($\gev$)}   & \multicolumn{2}{c|}{($\gev$)} &  $(\nb/\gev^2)$\\
\hline
\hline
\multirow{9}{*}{2$-$3} & \multirow{9}{*}{2.5 } &30 & $-$ & 50  & 39&5  &3.23  $\pm$ 0.40  $ ^{+0.43 }_{-0.33 }$ \\
                                           &   & 50 & $-$ & 70  &59&4  &5.35  $\pm$ 0.40  $ ^{+0.31 }_{-0.34 }$\\
                                           &   & 70 & $-$ & 80  &74&9  &7.13  $\pm$ 0.72  $ ^{+0.74 }_{-0.49 }$ \\
                                           &   & 80 & $-$ & 90  &84&9  &5.30  $\pm$ 0.62  $ ^{+0.47 }_{-0.50 }$ \\
                                           &   & 90 & $-$ & 100 &95&0  &6.04  $\pm$ 0.71  $ ^{+0.37 }_{-0.45 }$ \\
                                           &   &100 & $-$ & 110 &105&0  &6.28  $\pm$ 0.75  $ ^{+0.57 }_{-0.37 }$ \\
                                           &   &110 & $-$ & 120 &114&8  &6.42  $\pm$ 0.86  $ ^{+0.57 }_{-0.41 }$ \\
                                           &   &120 & $-$ & 130 &125&1  &7.52  $\pm$ 1.00  $ ^{+0.77 }_{-0.60 }$ \\
                                           &   &130 & $-$ & 160 &144&3  &6.80  $\pm$ 0.72  $ ^{+0.71 }_{-0.57 }$ \\
\hline
\multirow{9}{*}{3$-$5}&\multirow{9}{*}{3.8 } & 30 & $-$ & 50  & 39&7  & 3.10  $\pm$ 0.42  $ ^{+0.21 }_{-0.20 }$ \\
                                          &  & 50 & $-$ & 70  &59&6  & 4.23  $\pm$ 0.37  $ ^{+0.22 }_{-0.23 }$ \\
                                          &  & 70 & $-$ & 80  &74&9  & 5.28  $\pm$ 0.65  $ ^{+0.31 }_{-0.36 }$\\
                                          &  & 80 & $-$ & 90  &84&9  & 5.22  $\pm$ 0.67  $ ^{+0.48 }_{-0.45 }$\\
                                          &  & 90 & $-$ & 100 &94&9  & 4.29  $\pm$ 0.63  $ ^{+0.41 }_{-0.26 }$\\
                                          &  &100 & $-$ & 110 &105&0  & 6.42  $\pm$ 0.93  $ ^{+0.68 }_{-0.42 }$\\
                                          &  &110 & $-$ & 120 &115&0  & 5.79  $\pm$ 0.92  $ ^{+0.39 }_{-0.32 }$\\
                                          &  &120 & $-$ & 130 &125&0  & 6.55  $\pm$ 1.07  $ ^{+0.54 }_{-0.53 }$\\
                                          &  &130 & $-$ & 160 &143&9  &6.40  $\pm$ 0.75  $ ^{+0.77 }_{-0.63 }$ \\

\hline
\multirow{4}{*}{5$-$10}&\multirow{4}{*}{6.7 } & 30 & $-$ & 70  &53&9 &1.69  $\pm$ 0.20  $ ^{+0.14 }_{-0.14 }$\\
                                           &  & 70 & $-$ & 90  &79&9 &2.36  $\pm$ 0.28  $ ^{+0.20 }_{-0.25 }$\\
                                           &  & 90 & $-$ & 110 &99&9 &2.69  $\pm$ 0.35  $ ^{+0.18 }_{-0.23 }$\\
                                           &  &110 & $-$ & 160 & 133&8 &3.61  $\pm$ 0.36  $ ^{+0.34 }_{-0.28 }$\\
\hline
\multirow{3}{*}{10$-$20}&\multirow{3}{*}{13.3 } & 50 & $-$ & 80   & 63&9   & 0.50  $\pm$ 0.108 $ ^{+0.05 }_{-0.07 }$\\
                                             &  &80 & $-$ & 120  & 98&8   &0.72  $\pm$ 0.132 $ ^{+0.07 }_{-0.12 }$\\
                                             &  &120 & $-$ & 160 & 139&8  &0.84  $\pm$ 0.188 $ ^{+0.14 }_{-0.10 }$\\

\hline
\end{tabular}
\caption{The cross-section $\sigma(\gamma p \to J/\psi \ Y)$ as a function of $W$ in four 
$|t|$ bins and for $z>0.95$. The first uncertainty is statistical and
the second is systematic.}
\label{tab-table4}
\end{center}
\end{table}

%-------------------------

\begin{table}[p]
\begin{center}
\begin{tabular}{|rp{.1cm}r|cr@{.}lc|r@{ $ \pm $ }r|r@{ $ \pm $ }r|r@{ $ \pm $ }r|}
\hline
%\multicolumn{3}{|c|}{$|t|$ bin} &
%\multicolumn{4}{c|}{$\langle|t|\rangle$} &
%\multicolumn{2}{c|}{$r^{04}_{1-1}$} &
%\multicolumn{2}{c|}{$r^{04}_{00}$} &
%\multicolumn{2}{c|}{$Re\{{r^{04}_{10}}\}$}\\

\multicolumn{3}{|c|}{$|t|$ bin} &
\multicolumn{4}{c|}{ $\langle|t|\rangle$} & 
\multicolumn{2}{c|}{\multirow{2}{*}{$r^{04}_{1-1}$}} & 
\multicolumn{2}{c|}{\multirow{2}{*}{$r^{04}_{00}$}} &
\multicolumn{2}{c|}{\multirow{2}{*}{$Re\{{r^{04}_{10}}\}$}}\\
 
\multicolumn{3}{|c|}{($\gev^2$)} & 
\multicolumn{4}{c|}{($\gev^2$)}  &
\multicolumn{2}{c|}{} &
\multicolumn{2}{c|}{} &
\multicolumn{2}{c|}{} \\ 
%&\multicolumn{2}{c|}{a}&\multicolumn{2}{c|}{a} &\multicolumn{2}{c|}{a} & \\
\hline
\hline
2&$-$&3   && 2&5&  & 0.005    &0.064 $^{+0.019}_{-0.024}$   &   0.090  & 0.088 $^{+0.009}_{-0.017}$  &   0.117 & 0.061 $^{+0.025}_{-0.019}$  \\
3&$-$&5   && 3&8&  & $-$0.206 &0.072 $^{+0.021}_{-0.037}$   &$-$0.030  & 0.100 $^{+0.041}_{-0.083}$  &   0.197 & 0.068 $^{+0.046}_{-0.055}$  \\
5&$-$&10  && 6&7&  & 0.003    &0.106 $^{+0.036}_{-0.016}$   &$-$0.033  & 0.147 $^{+0.020}_{-0.042}$  &   0.154 & 0.088 $^{+0.016}_{-0.025}$  \\
10&$-$&20 &&13&3&  &$-$0.164  &0.240 $^{+0.108}_{-0.115}$   &$-$0.259  & 0.328 $^{+0.081}_{-0.062}$  &$-$0.153 &0.172 $^{+0.062}_{-0.047}$  \\
\hline
\end{tabular}
\caption{The spin density matrix elements for $30<W<160\gev$ and $z>0.95$. The first uncertainty is statistical and the second is systematic.}
\label{tab-table5}
\end{center}
\end{table}

%----------------------------------------

%------------------------------------------------------------------------------
%       Figures
%------------------------------------------------------------------------------
% figures......
%\newpage
% fig  1 --------------------------------
%
% fig  2 --------------------------------
%
\newpage
%fig. 3  ------------------------------------------------
\newpage
%fig 4 ----------------------
\newpage
% fig 5   --------------------------------------------
\newpage
%fig 6 -----------------------
\newpage
%fig 7 -----------------------
\newpage
%fig 8 ----------------------
\newpage
%fig 9 ----------------------
\newpage

% fig 10  --------------------------------------------

% fig 11  --------------------------------------------

%
%       ... that's it
%
\end{document}